\documentclass[twocolumn,aps,pra,longbibliography,superscriptaddress,nofootinbib]{revtex4-1}
\usepackage[utf8]{inputenc}
\usepackage{graphicx}
\usepackage{graphics}
\usepackage{amssymb}
\usepackage{color}
\usepackage{palatino}
\usepackage{nicefrac}
\usepackage{multibib}
\usepackage{bm}
\usepackage{mathrsfs}
\usepackage{amsmath}
\usepackage{amsfonts}
\usepackage{times,txfonts}
\usepackage{amsmath}
\usepackage{dsfont}
\usepackage{physics}
\usepackage{hyperref}
\usepackage{tikz}
\usepackage{float}
\usetikzlibrary{quantikz}
\usepackage{cancel}

\begin{document}

\title{Digital-analog quantum simulation of fermionic models}
\author{Lucas C. C\'{e}leri}
\affiliation{QPequi Group, Institute of Physics, Federal University of Goi\'{a}s, POBOX 131, 74001-970, Goi\^{a}nia, Brazil}
\affiliation{Department of Physical Chemistry, University of the Basque Country UPV/EHU, Apartado 644, 48080 Bilbao, Spain}

\author{Daniel Huerga}
\affiliation{Department of Physical Chemistry, University of the Basque Country UPV/EHU, Apartado 644, 48080 Bilbao, Spain}

\author{Francisco Albarr\'an-Arriagada}
\affiliation{International Center of Quantum Artificial Intelligence for Science and Technology (QuArtist) \\ and Department of Physics, Shanghai University, 200444 Shanghai, China}

\author{Enrique Solano}
\affiliation{Department of Physical Chemistry, University of the Basque Country UPV/EHU, Apartado 644, 48080 Bilbao, Spain}
\affiliation{International Center of Quantum Artificial Intelligence for Science and Technology (QuArtist) \\ and Department of Physics, Shanghai University, 200444 Shanghai, China}
\affiliation{IKERBASQUE, Basque Foundation for Science, Plaza Euskadi 5, 48009 Bilbao, Spain}
\affiliation{IQM, Nymphenburgerstr. 86, 80636 Munich, Germany}

\author{Mikel Garcia de Andoin}
\affiliation{Department of Physical Chemistry, University of the Basque Country UPV/EHU, Apartado 644, 48080 Bilbao, Spain}
\affiliation{TECNALIA, Basque Research and Technology Alliance (BRTA), 48160 Derio, Spain}
\affiliation{EHU Quantum Center, University of the Basque Country UPV/EHU, 48940 Leioa, Spain}

\author{Mikel Sanz}
\affiliation{Department of Physical Chemistry, University of the Basque Country UPV/EHU, Apartado 644, 48080 Bilbao, Spain}
\affiliation{EHU Quantum Center, University of the Basque Country UPV/EHU, 48940 Leioa, Spain}
\affiliation{IKERBASQUE, Basque Foundation for Science, Plaza Euskadi 5, 48009 Bilbao, Spain}
\affiliation{Basque Center for Applied Mathematics (BCAM), Alameda de Mazarredo 14, 48009 Bilbao, Spain}

\begin{abstract}
Simulating quantum many-body systems is a highly demanding task since the required resources grow exponentially with the dimension of the system. In the case of fermionic systems, this is even harder since nonlocal interactions emerge due to the antisymmetric character of the fermionic wave function. Here, we introduce a digital-analog quantum algorithm to simulate a wide class of fermionic Hamiltonians including the paradigmatic one-dimensional Fermi-Hubbard model. These digital-analog methods allow quantum algorithms to run beyond digital versions via an efficient use of coherence time. Furthermore, we exemplify our techniques with a low-connected architecture for realistic digital-analog implementations of specific fermionic models. 
\end{abstract}

\maketitle

\section{Introduction}

The use of quantum resources may allow us to improve a variety of classical tasks in computation~\cite{DiVincenzo1995}, communication~\cite{Ma2012}, and simulation~\cite{Feynman1982,Feynman1986}. In his seminal work, Feynman recognized that the complexity of simulating or computing quantum systems grows exponentially with the number of particles comprising the system~\cite{Feynman1982}. When the proposed solution is to employ another controllable quantum system to simulate the dynamics of the unknown one, we speak about analog quantum simulation. The latter has been successfully employed for paradigmatic cases such as the quantum Rabi model~\cite{Ballester2012,Braumuller2017,Lv2018}, the Dynamical Casimir effect~\cite{Felicetti2014,Rossatto2016,Sanz2018}, the Jaynes-Cummings and Rabi lattices~\cite{Hartmann2006,Hartmann2007,Greentree2006}, fermionic systems~\cite{Byrnes2008, Hensgens2017,ThesisLaura2017,Salfi2016,Tarruel2018}, as well as the recent boson sampling~\cite{Zhong2020}, just to name a few. Moreover, it is also possible to implement digital quantum simulations~\cite{Lloyd1996} with a number of interesting applications~\cite{Barends2015}.

Along these lines, quantum computing emerged with the formal proposal of a quantum Turing machine~\cite{De85,BV97}, the discovery of quantum algorithms with quantum speedup~\cite{Shor1994,Shor1996,Grover1997}, universal sets of quantum gates~\cite{Barenco1995}, and quantum error correction~\cite{Shor1995,Kitaev2003,Bennett1996}. This entire approach may be called digital quantum computing, given that it is based on an algorithmic sequence of single-qubit gates (SQG) and two-qubit gates~\cite{Deutsch1989}. Among key implementations of this paradigm in different quantum platforms, we can mention experiments in superconducting qubits~\cite{Barends2015,Barends2016,Klco2018,Aruteetal19} and ion traps~\cite{Lanyon2011,Martinez2016}.

Recently, an innovative quantum computing paradigm was proposed in Ref.~\cite{Parra-Rodriguez2020}, where digital-analog quantum computation (DAQC) was introduced. DAQC merges the digital methods, which provide versatility, with the analog approaches, which enhance robustness against errors, displaying better scalability than purely digital approaches in the same NISQ devices. This approach was used to propose the realistic implementation of the quantum Fourier transform~\cite{Martin2020} and the quantum approximate optimization algorithm (QAOA)~\cite{HMMSSW20}. Previously, a nonuniversal approach for digital-analog quantum simulations was developed and recently reviewed~\cite{Lamata2018}.

In this article, we develop the DAQC approach to simulate strongly-correlated fermionic systems by studying the paradigmatic Fermi-Hubbard (FH) model in one dimension (1D). This and related models are at the core of intense research due to its implications to high-temperature superconductivity, among other phenomena. Although its seemingly simple expression, an exact solution of the Hubbard model is only known for 1D~\cite{Lieb1968} and infinite dimensions~\cite{Metzner1989}. In dimensions of relevance for materials (2D, 3D), its simulation poses severe difficulties to state-of-the-art classical computational methods, such as the infamous sign-problem of quantum Monte Carlo~\cite{Troyer2005}, which has motivated tremendous efforts in the development of alternative numerical approaches~\cite{LeBlanc2015}. Our choice for the one-dimensional model is justified since our goal here is to show that the DAQC technique can be successfully employed in order to efficiently simulate a fermionic system. The extension of the techniques presented here to higher-dimensional systems is much more involved, including the required computational power to simulate the system on a classical computer, but can be performed. Of course, the architecture of the hardware will have to change, but according to our findings, we expect that the advantage of the digital-analog algorithm will overcome the purely digital one.

The difficulty of quantum simulating or computing fermionic models resides in the need of non-local gates to account for the antisymmetric character of the fermionic many-body wave function, as they appear when employing the Jordan-Wigner mapping~\cite{Jordan1928} relating fermions to qubits, the building blocks of quantum computers~\cite{Ortiz2001,Lanyon2010,Peruzzo2014,Barends2015,Hempel2018}. Then, by means of a Lie-Suzuki-Trotter decomposition~\cite{Trotter1959,Suzuki1976}, the unitary evolution generated by the qubit Hamiltonian can be written in terms of SQGs and two-qubit gates, making possible the digital computation of the fermionic Hamiltonian. On the analog side, the idea is to employ a suitable testbed system that mimics the fermionic dynamics, as employed in optical traps~\cite{OHara2002} and trapped ions~\cite{Kim2010}. Here, we employ tools from both of these approaches in order to show how the digital-analog paradigm can be used in the simulation of fermionic Hamiltonians. Due to the intrinsic digital-analog nature of our method, the implemented quantum algorithm will show a higher resilience against decoherence. We also find a low-connected architecture for optimal adaptation to current setups. 

In a recent work~\cite{Guseynov2022} a DAQC simulation a many body Fermionic system is proposed. There, they apply SWAP networks to systems with planar connectivity graphs, in particular square lattices, ladder topologies and linear graphs. For the latter, they obtain a circuit depth of $\mathcal{O}(n_q)$, where $n_q$ is the number of qubits employed. Recently a VQE algorithm was employed to simulate up to 16 qubits in a superconducting processor, employing error mitigation techniques on a short depth circuit~\cite{Stanisic2022}. Another recent work~\cite{Rubin2021} proposed a quantum inspired classical algorithm, in which they employ the symmetries of the system to reduce the complexity of the operations, pushing the boundaries of quantum advantage requirements for fermionic system simulations.

The Article is organized as follows. We start by describing our system in the next section, followed by the digital-analog proposal for the considered model, which naturally leads to optimal architectures. We illustrate the protocol by applying it to the one-dimensional Fermi-Hubbard Hamiltonian for a few fermions, including the action of noise. We close the manuscript with a summary of the results and a discussion on the physical and practical aspects of the proposed DAQC algorithm for fermionic models.

\section{The model} 

We will focus on simulating the one-dimensional Fermi-Hubbard (FH) model~\cite{Hubbard1963,Gutzwiller1964} on an $n$-site chain with open boundary conditions (OBC),
\begin{eqnarray}
H &=& \lambda\sum_{j,s}\left(c_{j,s}^{\dagger}c_{j+1,s} + c_{j+1,s}^{\dagger}c_{j,s}\right) +\epsilon\sum_{j}n_{j,\uparrow}n_{j,\downarrow}
\nonumber \\
&&+ \mu \sum_{j,s}n_{j,s} ,
\label{eq:hubbard}
\end{eqnarray}  
where $\lambda$ and $\epsilon$ describe the tunneling and the on-site interaction amplitudes, respectively, and $\mu$ is the chemical potential. The summation runs over all sites $1\le j\le = 1, ..., n$ of the system, and $s = \uparrow,\downarrow$ labels the spin. Thus, operator $c_{j,s}^{\dagger}$ creates a fermion at site $j$ with spin $s$, while $n_{j,s} = c_{j,s}^{\dagger}c_{j,s}$ denotes the number operator. Due to Pauli exclusion principle, these fermionic operators must fulfill the anticommutation relations $\lbrace c_{j,s},c_{l,s'}^{\dagger}\rbrace = \delta_{j,l}\delta_{s,s'}$ and $\lbrace c_{j,s}^{\dagger},c_{l,s'}^{\dagger}\rbrace = \lbrace c_{j,s},c_{l,s'}\rbrace = 0$. Such relations ensure global antisymmetry of the wave-function under exchange of fermions. 

\begin{figure}[t]
\includegraphics[width=0.45\textwidth]{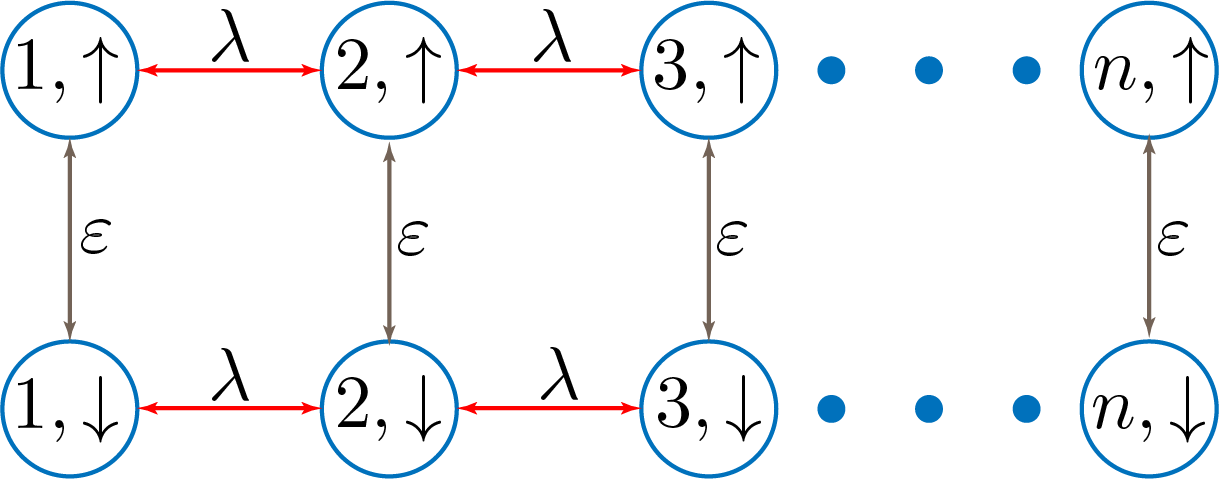}
\caption{\textbf{Qubit Hamiltonian.} The circles represent the qubits, each one labelled by a double index, the site in the lattice and the orientation of the spin. The arrows linking the circles represent interactions. Each chain contains $n$ qubits, the number of considered sites. The top chain holds the up state of the fermion while the bottom one is employed to represent the fermionic down state. The one site energy is not shown.}
\label{fig:spinqubitmap}
\end{figure}

In order to simulate the Hubbard dynamics in a quantum computer, we need to implement the unitary evolution $U_{H}^{t} = \mbox{exp}\lbrace -i H t\rbrace$. In general, quantum computers employ controlled qubits (which are neither fermions nor bosons) in order to implement a simulation of a given system. Therefore, the first goal is to map this fermionic dynamics into one describing a collection of qubits. We will do it by means of the Jordan-Wigner transformation~\cite{Jordan1928,Ortiz2001,Prosen2012,Reiner2016}. The idea is to associate the occupation number of a given fermionic mode ---which can be occupied or not--- with the two possible states of a qubit~\cite{Reiner2016}. Considering spin-less fermions, a creation fermionic operator maps to a \textit{string} of spin operators,
\begin{equation}
c_{j}^{\dagger} = \sigma_{j}^{+}\bigotimes_{i=1}^{j-1} \sigma^{z}_{i},
\label{eq:jw}
\end{equation}
where $c_j=(c_{j}^{\dagger})^\dag$ and $\sigma^{\pm}=(\sigma^x \pm i\sigma^y)/2$, and $\sigma^\alpha_j~(\alpha=x,y,z)$ are the Pauli matrices. It is straightforward to show that transformation~\eqref{eq:jw} preserves the fermionic anti-commutation relations and that  $c_{j}^{\dagger}c_{j+1} = \sigma_{j}^{+}\sigma_{j+1}^{-}$ and $c_{j}^{\dagger}c_{j} = \sigma_{j}^{+}\sigma_{j}^{-} = \left(\mathds{1} + \sigma_{j}^{z}\right)/2$. 

Now, considering a system of qubits with 1D topology, and that each site of the FH model~\eqref{eq:hubbard} has two fermionic modes corresponding to the spin, we assign the $s=\downarrow~(\uparrow)$ modes to the first (second) half of the qubit chain, i.e. $(j,\uparrow) \rightarrow j$ and $(j,\downarrow) \rightarrow j+n$. Therefore, we need $n_q = 2n$ qubits in order to simulate the 1D FH on $n$ sites. In this new notation, the FH Hamiltonian takes the form
\begin{eqnarray}
H = & \lambda & \sum_{j=1}^{n-1}\left(c_{j}^{\dagger}c_{j+1} + c_{j+1}^{\dagger}c_{j} + c_{j+n}^{\dagger}c_{j+n+1} + c_{j+n+1}^{\dagger}c_{j+n}\right) \nonumber \\ 
&+& \epsilon\sum_{j}^{n}c^{\dagger}_{j}c_{j}c^{\dagger}_{j+n}c_{j+n} + \mu \sum_{j}^{n}\left(c^{\dagger}_{j}c_{j} + c^{\dagger}_{j+n}c_{j+n}\right).
\end{eqnarray}
In terms of qubit/spin operators, this takes the expression
\begin{eqnarray}
H_{Q} = && \frac{1}{2}\left(\frac{\epsilon}{2} + \mu\right)  \sum_{j}^{n_q}\sigma_{j}^{z} + \frac{\epsilon}{4}\sum_{j}^{n}\sigma_{j}^{z}\sigma_{j+n}^{z} \nonumber \\ 
&& + \frac{\lambda}{2} \sum_{j=1}^{n - 1}\left(\sigma_{j}^{x}\sigma_{j+1}^{x} + \sigma_{j}^{y}\sigma_{j+1}^{y}\right) \nonumber \\
&& + \frac{\lambda}{2} \sum_{j=1}^{n - 1}\left(\sigma_{j+n}^{x}\sigma_{j+n+1}^{x} + \sigma_{j+n}^{y}\sigma_{j+n+1}^{y}\right).
\label{eq:qubit_hamiltonian}
\end{eqnarray}

As shown in Fig.~\ref{fig:spinqubitmap}, the 1D FH model naturally maps to a ladder qubit system where each qubit of rung $j$ in the left (right) leg encodes the spin-up (spin-down) fermionic mode of site $j$ in the 1D FH. In our case, we are assigning each leg of the ladder to each half of the qubit 1D hardware~\cite{Prosen2012,Reiner2016}. An interesting generalization of this transformation for two dimensions was recently proposed in Ref.~\cite{Steudtner2019}.

\section{Digital-analog quantum algorithm}

The digital-analog approach makes use of the native Hamiltonian in the platform system, together with single-qubit rotations, to implement the quantum computation. For simplicity, let us consider that the system in our control is governed by the Ising with nearest-neighbor interactions Hamiltonian
\begin{equation}\label{eq:ising}
H_{I} = \sum_{i=1}^{n_q - 1}\beta_i\sigma_{i}^{z}\sigma_{i+1}^{z},
\end{equation}
with $\beta_i$ a coupling constant. This comprises the analog part of the simulation. Therefore, this Hamiltonian, along with one-qubit gates, is our resource for building the simulation of the Hamiltonian~\eqref{eq:hubbard}. The central idea is to find a sequence of digital and analog blocks that maps the evolution under $H_{I}$ onto the evolution under $H_{Q}$~\eqref{eq:qubit_hamiltonian}.

\begin{figure}[t]
    \centering
    \includegraphics[width = \linewidth]{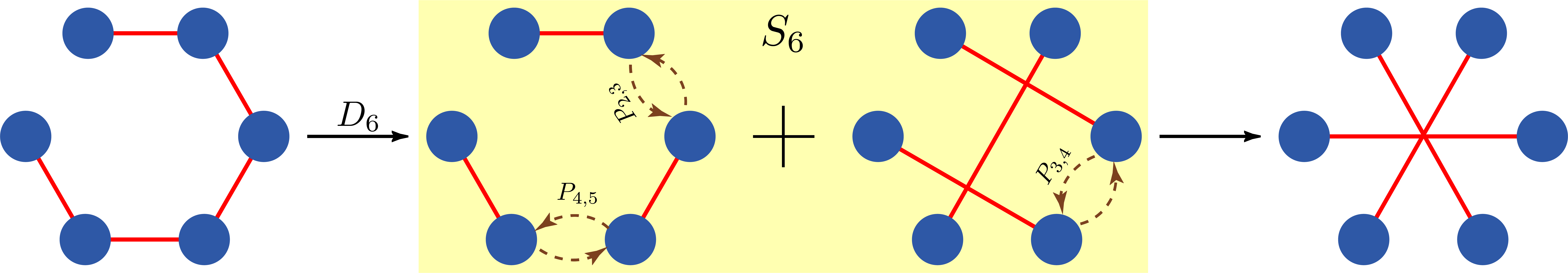} \\ \hspace{0.1cm} 
    \includegraphics[width = \linewidth]{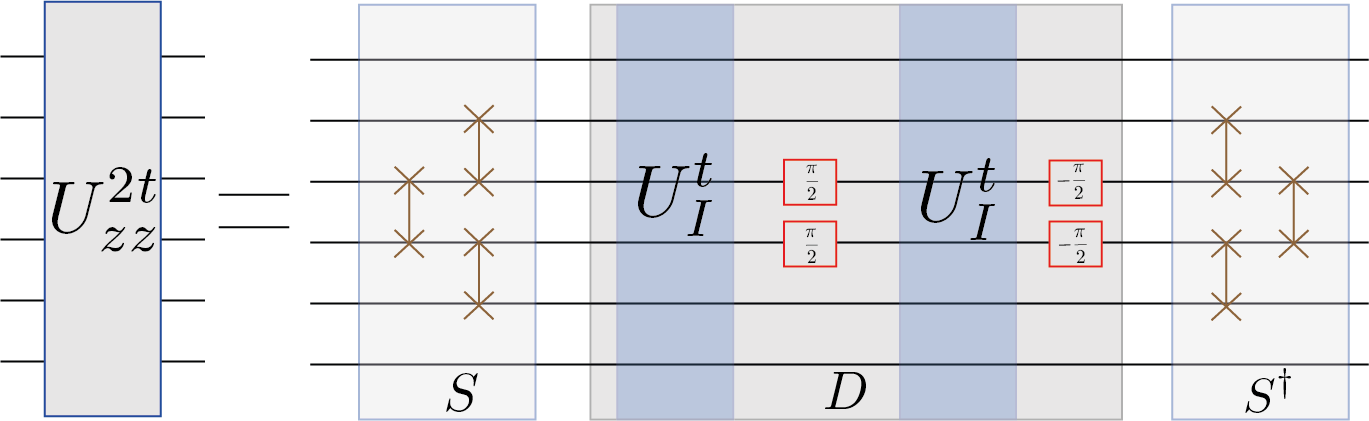}
    \caption{\textbf{Computation of $H_{ZZ}$}. The top panel shows the graph representation of the mapping $H_{I}\rightarrow H_{ZZ}$, given in Eq.~\ref{eq:hzz}, for 3-site Fermi-Hubbard. The first graph on the left represents the Ising Hamiltonian $H_{I}$ given in Eq.~\eqref{eq:ising} of the main text for the case $n_q = 6$. The blue spheres are the vertices while the red lines represents the interaction between two linked vertices. $D_6$ is the operation that decouples the necessary interactions. Transformation $S_6 = P_{3,4}P_{4,5}P_{2,3}$, represented by the two graphs in the shaded area of the figure, implements two sequences of {\sc SWAP} gates, with a total of 4 SWAP layers, as indicated by the dashed arrows in the figures. $P_{i,j}$ means the permutation of sites $i$ and $j$. The result is the desired graph shown on the right, that represents the Hamiltonian $H_{ZZ}$. The bottom panel shows the quantum circuit of the process.}
    \label{fig:Circuit3fzz}
\end{figure}

\begin{figure}[t]
\centering
    \includegraphics[width = 0.95\linewidth]{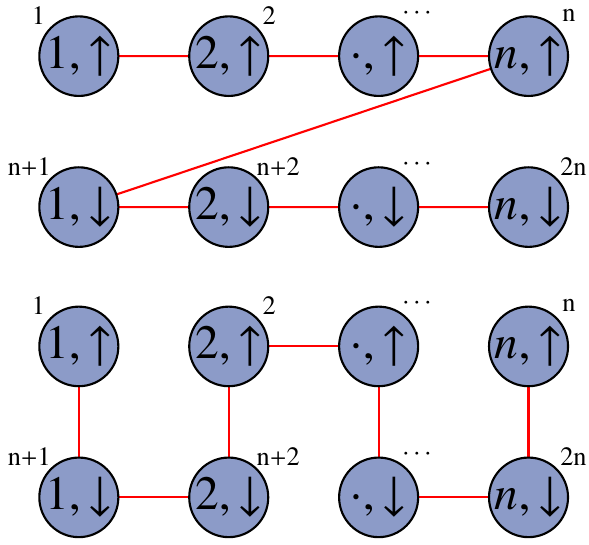}
    \caption{\textbf{Qubit mappings onto a linear hardware}. On top, the direct mapping of the qubits, and below, the ``snake-like" mapping. We depict in red the physical linear device employed for the simulation. The numeric labels correspond to the ones employed in equation~\eqref{eq:qubit_hamiltonian}. With the direct mapping, the maximum distance between the two qubits representing the states up and down for the same fermion is equal to the number of fermions $n$. On the contrary, the snake-like mapping reduces this distance to 3 for any number of fermions.}
    \label{fig:mappings}
\end{figure}

First, we must select a correct mapping of the qubits onto the hardware we are employing. As a first approach, one can directly assign the labels in Eq.~\eqref{eq:qubit_hamiltonian} to those in Eq.~\eqref{eq:ising} ($1, 2, \dots, n, n+1, \dots, n+n$). However, this leads to having to simulate couplings between qubits that are at a distance of $n$. Simulating these long range interactions require to implement $2(n-1)$ layers of {\sc SWAP} gates, as shown in Fig.~\ref{fig:Circuit3fzz}. For reducing at maximum the number of {\sc SWAP} layers, we propose an alternating mapping of the up and down fermionic states ($1, n+1, n+2, 2, 3, n+3, n+4, \dots$), as illustrated in Fig.\ref{fig:mappings}. This new mapping reduces the maximum distance for the couplings from $n$ to 3, such that we can implement them using just two layers of {\sc SWAP} gates. To see this, note that if we apply now a {\sc SWAP} gate between all $i$ and $i+1$ qubits for all $i$ odd qubits we obtain a new configuration in which we find the remaining couplings ($n+1, 1, 2, n+1, n+3, 3, 4, \dots$). As in the previous case, we need to apply these SWAP gates after we simulate the interactions so that we can recover the original qubit mapping.

Then, we employ the Lie-Suzuki-Trotter decomposition~\cite{Trotter1959,Suzuki1976} to write the total evolution operator of the qubit representation of the FH mode,Eq.~\eqref{eq:qubit_hamiltonian}, as
\begin{equation}
U^{t} = \lim_{l\rightarrow \infty}\left[U^t_{XX^a}U^t_{YY^a}U^t_{Z}U^t_{ZZ}U^t_{XX^b}U^t_{YY^b}\right]^{l},
\label{eq:trotter}
\end{equation} 
where we employed the notation $U_{a}^{t}\equiv\exp (-i\Delta_l^tH_a)$, $l=n_\text{T}$ is the number of Trotter steps, $\Delta_{l}^{t} = t/l$ and
\begin{subequations}
\begin{eqnarray}
H_{Z} &=& \frac{1}{2}\left(\frac{\epsilon}{2} + \mu\right)\sum_{j}^{2n}\sigma_{j}^{z} \label{eq:hz} \, , \\ 
H_{ZZ} &=& \frac{\epsilon}{4}\sum_{j}^{n}\sigma_{j}^{z}\sigma_{j+n}^{z} \label{eq:hzz} \, , \\
H_{XX^a} &=& \frac{\lambda}{2}\left(\sum_{j=\{\text{even}\}}^{n-1}\sigma_{j}^{x}\sigma_{j+1}^{x} +\sum_{j=\{\text{odd}\}}^{n-1}\sigma_{n+j}^{x}\sigma_{n+j+1}^{x}\right) \label{eq:hxxa} \, , \\
H_{XX^b} &=& \frac{\lambda}{2}\left(\sum_{j=\{\text{odd}\}}^{n-1}\sigma_{j}^{x}\sigma_{j+1}^{x} +\sum_{j=\{\text{even}\}}^{n-1}\sigma_{n+j}^{x}\sigma_{n+j+1}^{x}\right) \label{eq:hxxb} \, , \\
H_{YY^a} &=& \frac{\lambda}{2}\left(\sum_{j=\{\text{even}\}}^{n-1}\sigma_{j}^{y}\sigma_{j+1}^{y} +\sum_{j=\{\text{odd}\}}^{n-1}\sigma_{n+j}^{y}\sigma_{n+j+1}^{y}\right) \label{eq:hyya} \, , \\
H_{YY^b} &=& \frac{\lambda}{2}\left(\sum_{j=\{\text{odd}\}}^{n-1}\sigma_{j}^{y}\sigma_{j+1}^{y} +\sum_{j=\{\text{even}\}}^{n-1}\sigma_{n+j}^{y}\sigma_{n+j+1}^{y}\right) \, . \label{eq:hyyb} 
\end{eqnarray}
\end{subequations}
This selection of the division of of the $H_{XX}$ and the $H_{YY}$ terms will lead to a reduction in the depth of the circuit, which will become clear in the following.

\begin{figure*}
    \centering
    \includegraphics[width = 0.95\linewidth]{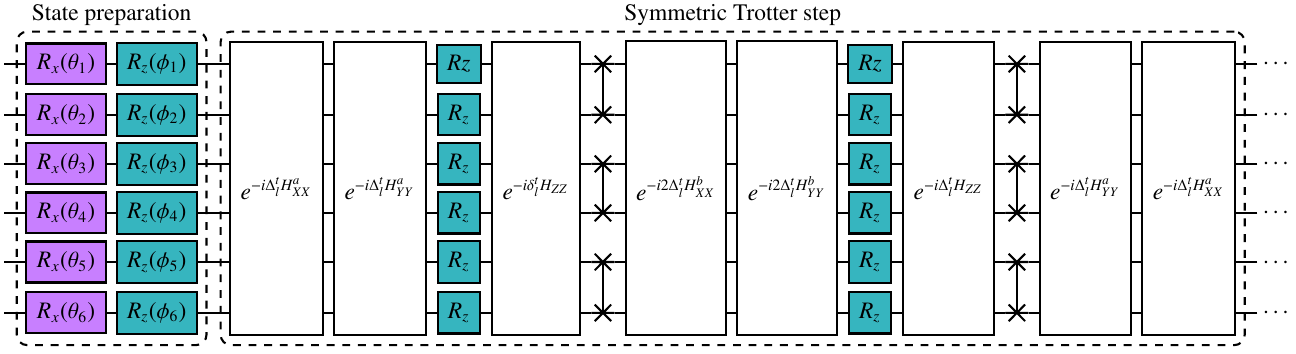}
    \caption{\textbf{Trotterized evolution}. Digital quantum circuit for implementing the Trotterized evolution under the fermionic Hamiltonian. Notice that the position of the qubits is the same at the start and at the end of each of the symmetric Trotter steps. We do not show here the SQGs to rotate the $\sigma^x\sigma^x$ and $\sigma^y\sigma^y$ interactions to the $\sigma^z\sigma^z$ axis.}
    \label{fig:digital}
\end{figure*}

The implementation of the schedule proposed in Eq.~\eqref{eq:trotter} requires two layers of {\sc SWAP} gates per Trotter step. With the objective of reducing even more the number of {\sc SWAP} layers, let us write a symmetric Trotterization~\cite{Suzuki1985} employing the same number of exponential terms
as in Eq.~\eqref{eq:trotter}
\begin{equation}
\label{eq:trotterSymmetric}
\begin{split}
U^{t}_\text{s} = \lim_{l\rightarrow \infty}&\left[U_{XX^a}^tU_{YY^a}^tU_{Z}^tU_{ZZ}^tU_{XX^b}^tU_{YY^b}^t\right.\\
&\phantom{[}\left.U_{YY^b}^tU_{XX^b}^tU_{ZZ}^tU_{Z}^tU_{YY^a}^tU_{XX^a}^t\right]^{l/2}.
\end{split}
\end{equation}
Each of these new Trotter steps requires again two {\sc SWAP} layers, as shown in Fig.\ref{fig:digital}. However, we have cut in half the number of {\sc SWAP} layers for the total circuit. Additionally, the use of the symmetric Trotter formula reduces the error from $\mathcal{O}(l^{-2})$ to $\mathcal{O}(l^{-3})$ while employing a circuit with the practically same number of exponential terms. Furthermore, notice that due to their commutation properties, we can merge all adjacent $U^t_{XX}$ and $U^t_{YY}$ terms, even those belonging to different Trotter steps.

\begin{figure}[h]
    \raggedright
    a)\\ \includegraphics[width=0.95\linewidth]{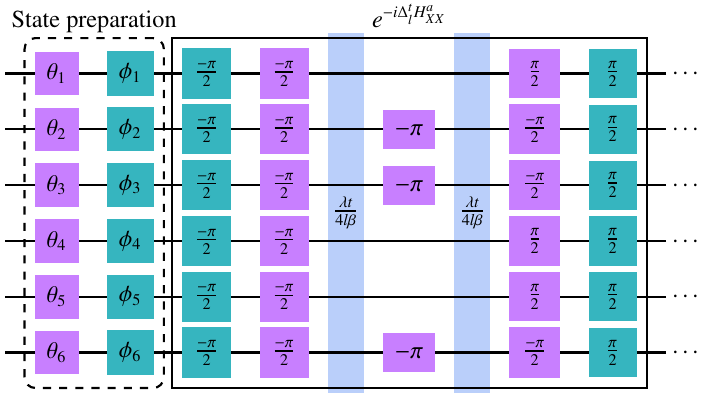}
    
    b)\\ \includegraphics[width=0.95\linewidth]{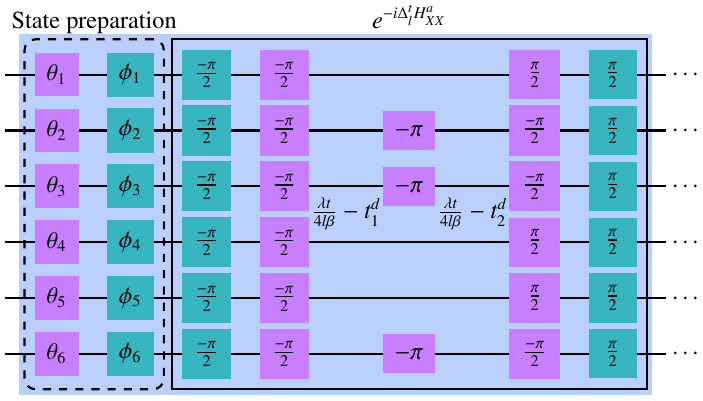}
    \caption{\textbf{Digital-analog schedule}. (a) sDAQC schedule for the state preparation and the first term from the first Trotter step (see Fig.\ref{fig:digital}). See here that in order to simulate the evolution under the $H_{XX}^a$ Hamiltonian we need to apply a $Y^{\pm\pi/2}$ rotation to all qubits before and after the analog blocks. For this, we are assuming that our hardware only allows us to implement $x$ and $z$ rotations, so we have employed the following identity, $Y^\theta=Z^{\pi/2}X^\theta Z^{-\pi/2}$. Employing the fact that single qubit $z$ rotations commute with the evolution under the Ising Hamiltonian and by merging all the adjacent SQGs of the same type, we have simplified the circuit to as shown in the figure. In blue we depict the hardware Ising Hamiltonian, in purple and green the single qubit $x$ and $z$ gates respectively. In the sDAQC circuit, the interaction Hamiltonian is turned off during the application of the SQGs. The angles for the SQGs are shown inside the squares, while the time for the analog blocks is shown between them. (b) Corresponding bDAQC schedule. Note that since the Hamiltonian is always on, the length of the analog blocks is reduced to fit the SQGs while keeping the time in which the Hamiltonian is on, with $t^d_1=5/2t_x+2t_z$ and $t^d_2=3/2t_x+t_z$.}
    \label{fig:sDAQCbDAQC}
\end{figure}

Now we have to design a digital-analog schedule for simulating these terms. For simplicity, let us start by sketching the digital circuit that implements the evolution in Eq.~\eqref{eq:trotterSymmetric}, as shown in Fig.\ref{fig:digital}. The $H_Z$ term from Eq.~\eqref{eq:hz} can be directly implemented with single qubit $z$ rotations. In order to implement the rest of interacting terms, we employ various set of two-qubit and single qubit gates (SQG). In this work instead, we will employ a digital-analog schedule, where we employ the natural interaction Hamiltonian from the hardware to implement the interactions. The digital-analog schedule is the sequence of SQGs (digital blocks) sandwiching control-free evolution periods (analog blocks). We distinguish two versions of DAQC, the stepwise (sDAQC) and the banged version (bDAQC). The difference between both is in the way the digital blocks are applied. In sDAQC, the interaction Hamiltonian is switched off during the digital block. On the contrary, in bDAQC, the driving for applying the SQGs is applied on top of the interaction. Assuming square pulse drivings, the description of a single analog block sandwiched between two digital blocks is
\begin{equation}
\begin{split}
    &U_\text{sDAQC}=e^{-it_\text{SQG}H_\text{SQG,1}}e^{-itH_I}e^{-it_\text{SQG}H_\text{SQG,2}}\\
    \approx&U_\text{bDAQC}=e^{-it_\text{SQG}(H_\text{SQG,1}+H_I)}e^{-i(t-2t_\text{SQG})H_I}e^{-it_\text{SQG}(H_\text{SQG,2}-H_I)},
\end{split}    
\end{equation}
where $H_\text{SQG}$ is the driving Hamiltonian for implementing the SQGs and $t_\text{SQG}$ the duration of the driving, such that the SQG can be written as $U_\text{SQG}=e^{-it_\text{SQG}H_\text{SQG}}$,. This difference introduces a small deviation in the unitary evolution of bDAQC compared to sDAQC, which gets reduced as the time to apply the SQGs decreases. Even though the performance of the sDAQC circuits will give a better fidelity compared to bDAQC, it might be challenging to turn on and off the interaction Hamiltonian in actual experiments. Thus, the bDAQC model provides a realistically implementable description of the DAQC paradigm. The comparison between both is further depicted in Fig.~\ref{fig:sDAQCbDAQC}.

While we can implement the {\sc SWAP} gate in a digital circuit employing 3 {\sc CNOT}s, for the digital-analog implementation we decompose the {\sc SWAP} gate between the qubits $i$ and $j$ into interactions in the Pauli basis, SWAP$_{ij}$ = $\exp[-i (\pi/4) (\sigma_i^x\sigma_j^x + \sigma_i^y\sigma_j^y+\sigma_i^z\sigma_j^z)]$. As we have done in a previous step, when writing all the terms in the Pauli basis we can merge adjacent terms with the same type of interactions.

The next step is to simulate the Hamiltonians with $\sigma_k^x \sigma_{k+1}^x$ and $\sigma_k^y \sigma_{k+1}^y$ with the resources from the hardware Hamiltonian $H_I$. For this, we rotate the qubits to the desired axis. Let us employ a compact notation for the SQGs, $\alpha_k^\theta\equiv R_{\alpha}^{k}(\theta) = \exp(-i \sigma_k^\alpha\theta/2)$, $\alpha=\{x,y,z\}$. Noticing that $\exp(-i t \sigma_k^x\sigma_{k+1}^x)=Y_1^{-\pi/2}Y_2^{-\pi/2}\exp(-i t \sigma_k^z\sigma_{k+1}^z)Y_1^{\pi/2}Y_2^{\pi/2}$ and $\exp(-i t \sigma_k^y\sigma_{k+1}^y)=X_1^{\pi/2}X_2^{\pi/2}\exp(-i t \sigma_k^z\sigma_{k+1}^z)X_1^{-\pi/2}X_2^{-\pi/2}$, we can express each Trotter step in Eq.~\eqref{eq:trotterSymmetric} in terms of SQGs and interaction terms between neighbouring qubits of the form $\exp(-it \beta_i\sigma_i^z\sigma_{i+1}^z)$. 

With all terms written as $\sigma_i^z\sigma_{i+1}^z$ interactions, we can implement them in a quantum hardware whose coupling connectivity completely covers the target couplings we want to simulate. The simulation can be implemented by applying a sequence of $X^{\pm\pi}$ rotations (digital blocks) and control-free evolutions (analog blocks). In Ref.\cite{Parra-Rodriguez2020} it was shown how to generate a digital-analog schedule in order to simulate an inhomogeneous Ising Hamiltonian employing another inhomogeneous Ising Hamiltonian as a resource. 

As an example, let us consider a simpler case. Let us assume that we want to simulate the evolution of a toy inhomogeneous nearest-neighbor Ising Hamiltonian $H_\text{toy}=\sum_{i=1}^{n-1}\alpha_i\sigma_i^z\sigma_{i+1}^z$ for a time $t_f$. As a resource, lets assume we have an inhomogeneous Ising Hamiltonian as in Eq.~\eqref{eq:ising}. Using the fact that $X^\pi\sigma^{z}X^{-\pi} = X^{-\pi}\sigma^{z}X^\pi = -\sigma^{z}$ we can invert the sign of a coupling by applying an $X$ gate on one of the two qubits. In this way, we can write
\begin{equation}
\begin{split}
    U_\text{toy}^{t_f}&=\exp\left(-i t_f \sum_{i=1}^{n-1}\alpha_i\sigma_i^z\sigma_{i+1}^z\right)\\
    &=\prod_{k=1}^{n-1}\left[\left(\prod_{i=1}^{n} X_k^{\pi A_{k,i}}\right)\exp\left(-it_kH_\text{I}\right)\left(\prod_{i=1}^{n} X_k^{-\pi A_{k,i}}\right)\right]\\ 
    &=\exp\left(-i\sum_{j=1}^{n-1}\sum_{k=1}^n (-1)^{A_{k,j}+A_{k,j+1}}t_k\beta_i\sigma_j^z\sigma_{j+1}^z\right),
\end{split}
\end{equation}
where $A_{k,i}\in\{0,1\}$ is a matrix that marks the application of $X^{\pm\pi}$ gates on qubit $i$ at the start and end on the $k^\text{th}$ analog block. Now, we have to solve the following system of equations
\begin{equation}
    t_f\alpha_i=\sum_{k=1}^{n-1}(-1)^{A_{k,i}+A_{k,i+1}}t_k\beta_i,\ \forall i\in\{1,\dots,n_q\}.
\end{equation}
We are free to choose any $A$ matrix which generates a compatible system of equations. Following the same reasoning as in ~\cite{Parra-Rodriguez2020}, we can choose to invert just one of the couplings per analog block. Then, an easy choice would be to apply the digital gates to the first $k$ qubits, this is $A_{k,i}=1$ if $i\leq k$ and $A_{k,i}=0$ else. With this, we can directly solve the equations to obtain the corresponding analog block times $\{t_k\}$.

From this general example, let us now go to the problem of generating the digital-analog schedule for our simulation. In this case, we will have to simulate the following nearest-neighbour Hamiltonians with alternating couplings
\begin{equation}
    H_\text{sim}=\nu\sum_{i=\{\text{odd}\}}^{n-1}\sigma_i^z\sigma_{i+1}^z + \xi\sum_{i=\{\text{even}\}}^{n-1}\sigma_i^z\sigma_{i+1}^z,
\end{equation}
with $\nu$ and $\xi$ the corresponding couplings. For simplicity, let us assume that we will employ a system whose couplings are designed to be equal, so we have $\beta_i=\beta\ \forall i$ in Eq.~\eqref{eq:ising}. When solving this system, we notice that it can be reduced to just a set of two equations if we employ the following selection of $X^{\pm\pi}$ gates, $A_{1,i}=1\ \forall i$, $A_{2,j}=1\ j\in\{1,4,5,8,9,\dots,2n-1\}$ if $\xi>\nu$, $A_{2,j}=1\ j\in\{2,3,6,7,10,\dots,2n\}$ if $\xi<\nu$, and $A_{2,j}=0$ else. This way, the time duration of the analog blocks is $t_1=(\xi+\nu)/2$ and $t_2=|\xi+\nu|/2$. 

With all these ingredients, we can now write the full digital-analog circuit for simulating the evolution. By applying every optimization trick mentioned in this section, the total time for the digital-analog circuit is
\begin{multline}
    T_\text{total}=3t_z+7t_x+l\left[4t_z+12t_x+t_{ZZ}\right.\\\left.+t_{\pi/4}+t_{XX}+\text{max}\left(t_{\pi/4},t_{XX}\right)\right],
\end{multline}
where $t_x$ ($t_z$) is the time to implement a digital $x$-($z$-)rotation gate, $t_{ZZ}=(\pi+(t\epsilon/l))/(4\beta)$, $t_{\pi/4}=\pi/(4\beta)$, and $t_{XX}=\lambda t/(l\beta)$. In total, we have to apply $3+8l$ analog blocks, $2+4l$ $Rz$ digital blocks, and $6+12l$ $Rx$ digital blocks. As we see, the digital-analog implementation scales linearly with the number of Trotter steps, and the total simulation time, $\mathcal{O}(tl)$.

\subsection{Ladder architecture}

In the previous section, it was assumed that the connectivity of the qubits in the hardware is linear, as shown in the upper panel of Fig.~\ref{fig:Circuit3fzz}, with the underlying Hamiltonian given by the Ising model in Eq.~\eqref{eq:ising}. This forced us to apply a set of {\sc SWAP} gates in order to implement the $H_{ZZ}$ Hamiltonian from Eq.~\eqref{eq:hzz}, since it contains long-range interactions. This introduces an extra cost in time that might have an impact on the performance of the circuit. We can avoid this problem by employing a more appropriate chip architecture.

\begin{figure}[t]
    \centering
    \includegraphics[width = 0.45\textwidth]{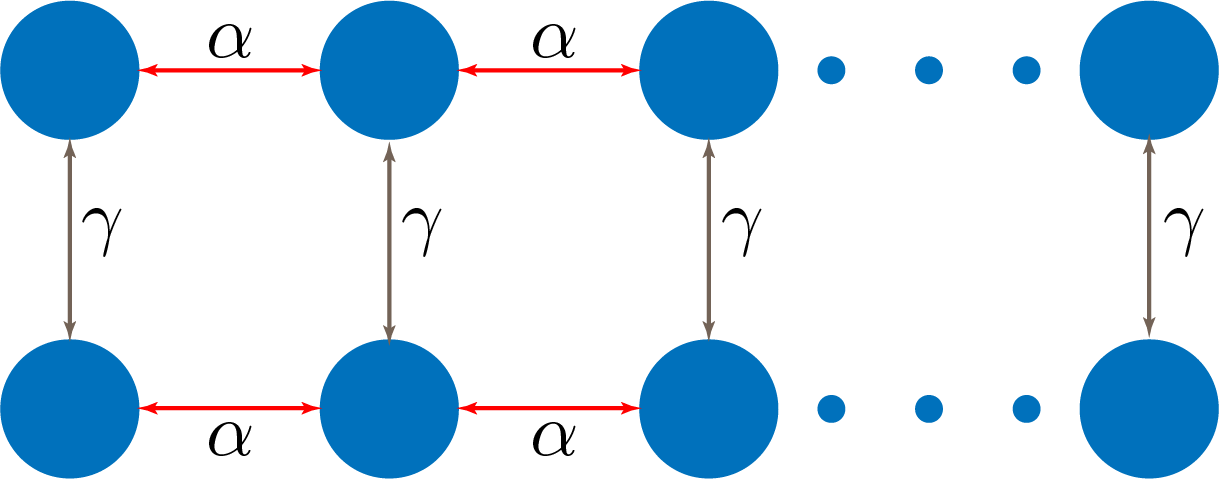}
    \caption{\textbf{Ladder architecture}. The blue circles represent the physical qubits, which interact with each other with strengths $\alpha$ and $\gamma$ with a $ZZ$ interaction. Given such ladder architecture, in order to implement the $H_{ZZ}$ Hamiltonian (Eq.~\eqref{eq:hzz}) all we need to do is to decouple all the $\alpha$ interactions. This can be done by single qubit rotations as explained in the text. To implement the $H_{XX}$ (Eqs.\ref{eq:hxxa},\ref{eq:hxxb}) and $H_{YY}$ (Eqs.\ref{eq:hyya},\ref{eq:hyyb}) interactions, we need to decouple all the qubits linked by $\gamma$. This can also be done by single qubit rotations. Moreover, by employing such single qubit rotations, we can change the coupling between the desired qubits in order to achieve the target ones given by $\epsilon$ and $\lambda$, in the same way we did in the examples discussed in the text.}
    \label{fig:optimal}
\end{figure}

This means that we can start from a different architecture, in which the underlying Hamiltonian has already the same connectivity as the problem, thus eliminating the necessity of all the {\sc SWAP} gates. This new architecture would decrease the overall error and time of the computation, thus decreasing the number of necessary Trotter steps to achieve a given precision. Figure~\ref{fig:optimal} shows the optimal architecture in the sense of the one which would require the minimal number of digital and analog blocks. This processor architecture emerges naturally from the Hamiltonian in Eq.~\eqref{eq:qubit_hamiltonian},
\begin{equation}
    H_\text{ladder}=\sum_{i=1}^{n-1}\alpha_{i}\sigma_i^z\sigma_{i+1}^z+\alpha_{n+i}\sigma_{n+i}^z\sigma_{n+i+1}^z+\sum_{i=1}^n\gamma_i\sigma_i^z\sigma_{n+i}^z,
\end{equation}
with $\alpha_i$ and $\gamma_i$ the corresponding couplings.

With this new co-designed processor we can directly implement the FH model by employing a reduced number of steps compared to Eq.~\eqref{eq:trotterSymmetric}. Again, employing a symmetric Trotter formula, we can write a single step as
\begin{equation}
    U_s^t=\lim_{l\rightarrow\infty}\left[U_{XX}^{t/2}U_{YY}^{t/2}U_Z^tU_{ZZ}^tU_{YY}^{t/2}U_{XX}^{t/2}\right]^l,
\end{equation}
where we have already merged all commuting terms.

Now, for implementing each of the interacting terms, we have to first rotate the qubits as before. However, now there is a difference in the digital-analog blocks. Now, we have to simulate the evolution under Hamiltonians with the form
\begin{equation}
    H_\text{ladder}=\nu\sum_{i=1}^{n-1}\left(\sigma_i^z\sigma_{i+1}^z+\sigma_{n+i}^z\sigma_{n+i+1}^z\right)+\xi\sum_{i=1}^n\sigma_i^z\sigma_{n+i}^z,
\end{equation}
with $\nu$ and $\xi$ the corresponding coupling constants. Assuming that we have enough control over the design of the system, we can set the couplings so that all horizontal couplings are equal, and so with all vertical couplings, this is $\alpha_i=\alpha$ and $\gamma_i=\gamma\ \forall i$. This selection allows us to simulate each ladder Hamiltonian with just two analog blocks with time duration $t_1=[(\xi/\gamma)+(\nu/\alpha)]/2$ and $t_2=\lvert(\xi/\gamma)-(\nu/\alpha)\rvert/2$. A design of the digital blocks that allows for this distinguises two cases. If we have to simulate a $H_\text{ladder}$ with $\xi/\gamma>\nu/\alpha$ we apply $A_{1,i}=1\ \forall i,\ A_{2,j}=1$ for $j=\{1,3,5,\dots,n+1,n+3,n+5,\dots\}$, $A_{3,k}=1$ for $k=\{2,4,6,\dots,n+2,n+4,n+6,\dots\}$, and $A=0$ else. On the contrary, when $\nu/\alpha>\xi/\gamma$, the digital blocks are $A_{1,i}=1\ \forall i,\ A_{2,j}=1$ for $j=\{1,2,\dots,n\}$, $A_{3,k}=1$ for $k=\{n+1,n+2,\dots,2n\}$, and $A=0$ else.

\subsection{Hardware requirements for DAQC}

The digital-analog paradigm was introduced as a suitable quantum computing platform in the current NISQ era~\cite{JianWei2022}. One of the main advantages is the absence of two-qubit gates in the circuits, which allows us to evade the difficult task of having to reduce the gates infidelities to extract meaningful results from the experiments. In this sense, a suitable platform for DAQC could be any in which we can address each qubit so that we can implement arbitrary single qubit rotations, and in which the system Hamiltonian has a global entangling dynamics~\cite{Parra-Rodriguez2020}.

A well established platform for performing quantum computational task is the trapped ions setup~\cite{Cirac1995}. In the Lamb-Dike regime, a chain of trapped ions can be modeled with the Ising Hamiltonian. This, directly allows us to implement the DAQC circuits, in which the single qubit gates can be engineered through Stark shifts mediated by lasers~\cite{Staanum2002}.

Superconducting circuits are one of the most studied platforms in DQC. In this circuits, we can employ different circuit designs to build the qubits (e.g. transmon~\cite{Koch2007} or flux qubits~\cite{Yan2016}) and their couplings (direct capacitive~\cite{Pashkin2003} or inductive couplings~\cite{Kafri2017}, or resonator mediated couplings~\cite{Yan2016,Kafri2017,Retamal2022}). The Hamiltonian arising from these circuits can be then manipulated by changing the frame of reference in order to match the desired couplings~\cite{Rigetti2010, Tasio2021}. DAQC algorithms have already been implemented in superconducting circuits, showing their capabilities as a valid platform for this paradigm.

We argue that the current technology is ready for the simulation of the Fermi-Hubbard model. In Ref.~\cite{JianWei2022} they employed a circuit architecture with the qubits connected as a square lattice, which perfectly matches the proposed co-designed topology. The coupling strength they achieved in their system was in the order of $\sim$MHz. If we assume that the maximum coupling strength of the fermionic system ($\epsilon/2, \lambda/2$ from Eq.~\eqref{eq:hubbard}) is in the order of magnitude of $\sim$KHz (e.g.~\cite{Esslinger2010}), then we need to run the circuit for a time in the $\sim\mu$s. Taking IBM's systems as a example of the usual decoherence times of current superconducting circuits, $T_1\sim100\mu s$, we estimate that our algorithm could run in a real experiment.

\section{Numerical results}

In this section, we numerically study the digital-analog approach by inspecting the evolution of various observables and comparing them to the exact evolution.

In particular, we consider the total density
\begin{equation}
n_{i} = n_{i,\uparrow} + n_{i,\downarrow} = \mathds{1} + \frac{1}{2}\left(\sigma_{i}^{z} + \sigma_{i+n}^{z}\right),
\label{eq:density}
\end{equation}
and the on-site double-occupancy
\begin{equation}
n_{i,\uparrow}n_{i,\downarrow} = \frac{1}{4}\left(\sigma_{i}^{z}\sigma_{i+n}^{z} + 2n_{i}  - \mathds{1}\right),
\label{eq:correlations}
\end{equation}
which measures the on-site fermionic correlations and is customary used as a qualitative parameter related to the Mott insulator-metal transition.

In the following we will consider the time evolution of the average of these observables with respect to a randomly chosen separable initial state $\vert\psi\rangle = \vert\psi_{1}(\theta_{1},\phi_{1})\rangle\otimes\cdot\cdot\cdot\vert\psi_{n}(\theta_{n},\phi_{n})\rangle$, with $\vert\psi_{k}(\theta_{k},\phi_{k})\rangle = \cos\theta_{k}\vert 0\rangle_{k} + e^{-i\phi_{k}}\sin\theta_{k}\vert 1_{k}\rangle$ being the state of qubit $k$. The angles are randomly chosen in the intervals $\{\theta_{k},\phi_k\}\in [0,2\pi]$, such that we uniformly sample the Hilbert space of separable pure states.

To study the behavior of our digital-analog approach, we show a comparison between the time evolution of the analog (A), and both the banged (bDAQC) and the stepwise digital-analog (sDAQC) approaches for the 3-site fermion chain. In order to accelerate the numerical simulations we employ a simplified model for the DAQC circuits, we assume that all the control pulses applied are square pulses. This implies that for the simulation of sDAQC circuits, we simulate the switching of the interaction Hamiltonian is performed instantly. For the simulation of the analog schedule we calculate the exact analog evolution under the qubit Hamiltonian (Eq.~\eqref{eq:qubit_hamiltonian}). For calculating the fidelity, we have employed the usual formula
\begin{equation}
    F=\left\lvert\bra{\psi}U_\text{A}^\dagger\rho_\text{DA}U_\text{A}\ket{\psi}\right\rvert,
\end{equation}
where $U_{A}$ is the unitary evolution generated by the qubit Hamiltonian $H_{Q}$ (Eq.~\eqref{eq:hubbard}), and $\rho_\text{DA}$ is the density matrix representing the final state after the digital-analog circuit. 

\begin{figure}
    \raggedright
    a)
    \includegraphics[width=\linewidth,clip,trim={1.1cm 0.4cm 3.3cm 1.4cm}]{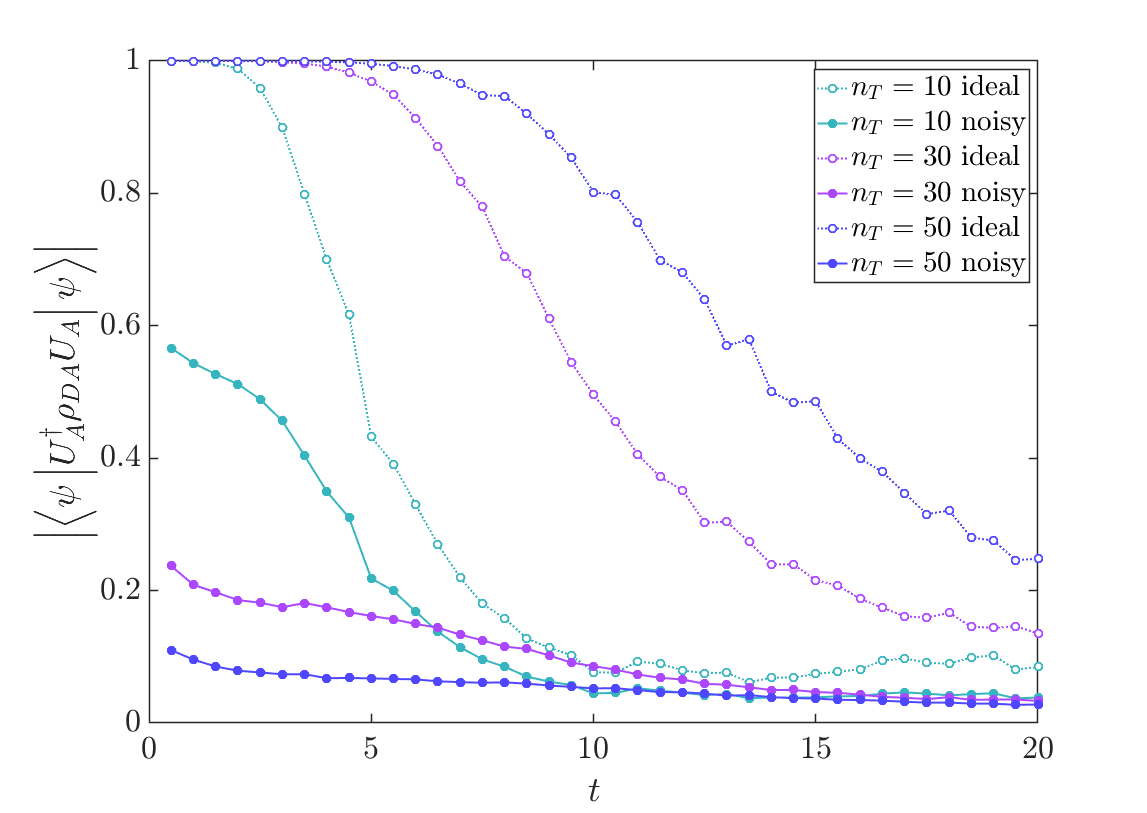}
    b)
    \includegraphics[width=\linewidth,clip,trim={1.1cm 0.4cm 3.3cm 1.4cm}]{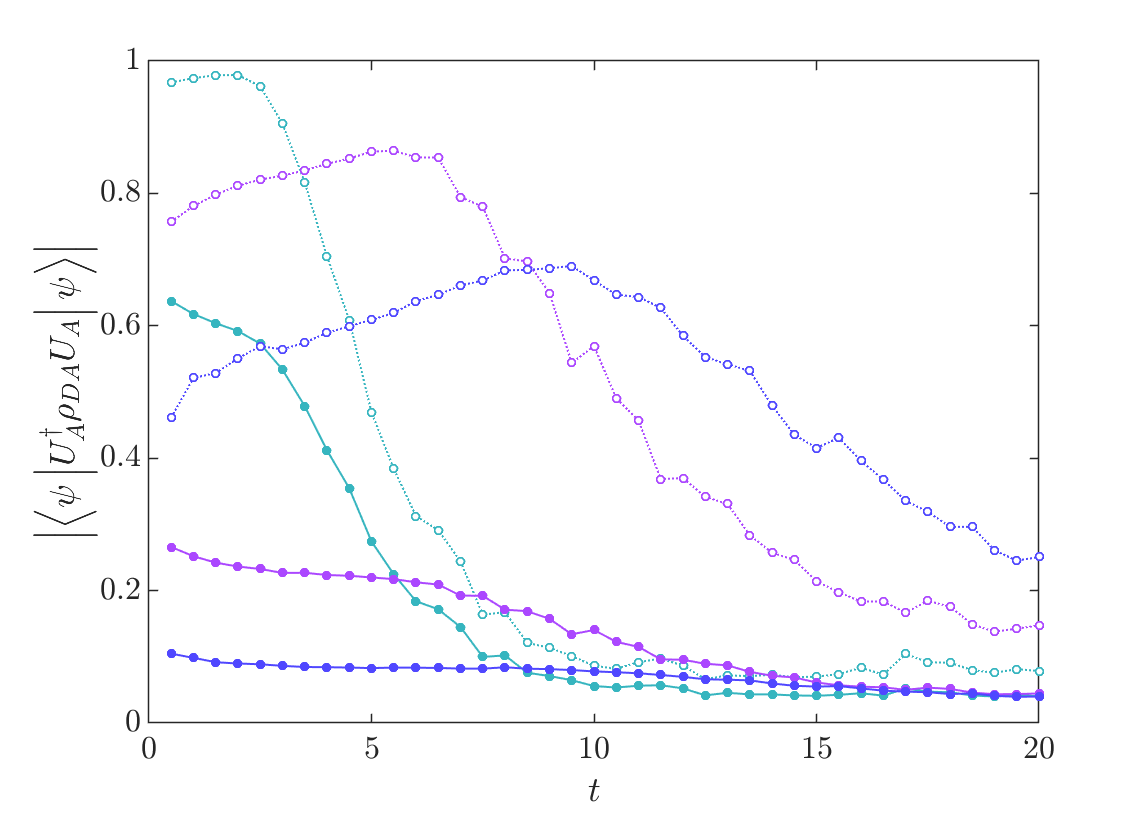}
    \caption{\textbf{Fidelity for the ``snake-like" mapping}. Fidelities for different number of Trotter steps and times for the (a) stepwise and (b) banged DA circuits. The dotted lines represent the ideal implementation of the circuits, while the solid ones shows the effect of the noise. Here we show the mean value over a maximum of 2000 runs with different initial random states. The fermionic Hamiltonian parameters are randomly set for each run with a uniform distribution in the ranges $\epsilon = [1/2,2]$, $\mu = [1/4,1]$, and $\lambda = [1/2,2]$, while the hardware Hamiltonian is set at $\beta = 1$. The time duration of the single qubit gates are $t_\text{SQG}=10^{-3}$. The error parameters are $r_\text{D} = 0.025$, $r_\text{B}^\text{(sDAQC)} = 20$, $r_\text{B}^\text{(bDAQC)} = 10^{-4}$, $p_\text{bf} = 10^{-4}$, $p_\text{th} = 0.35$, $T_1 = 100$, and $T_2 = 100$  (see Sec.~\ref{sec:error} for definitions). All parameters are given in units of $\hbar=1$, and represent realistic parameters for the current NISQ era superconducting circuits~\cite{Paula2021}.}
    \label{fig:snakeFidelity}
\end{figure}

\begin{figure}
    \raggedright
    a)
    \includegraphics[width=\linewidth,clip,trim={1.1cm 0.4cm 3.3cm 1.4cm}]{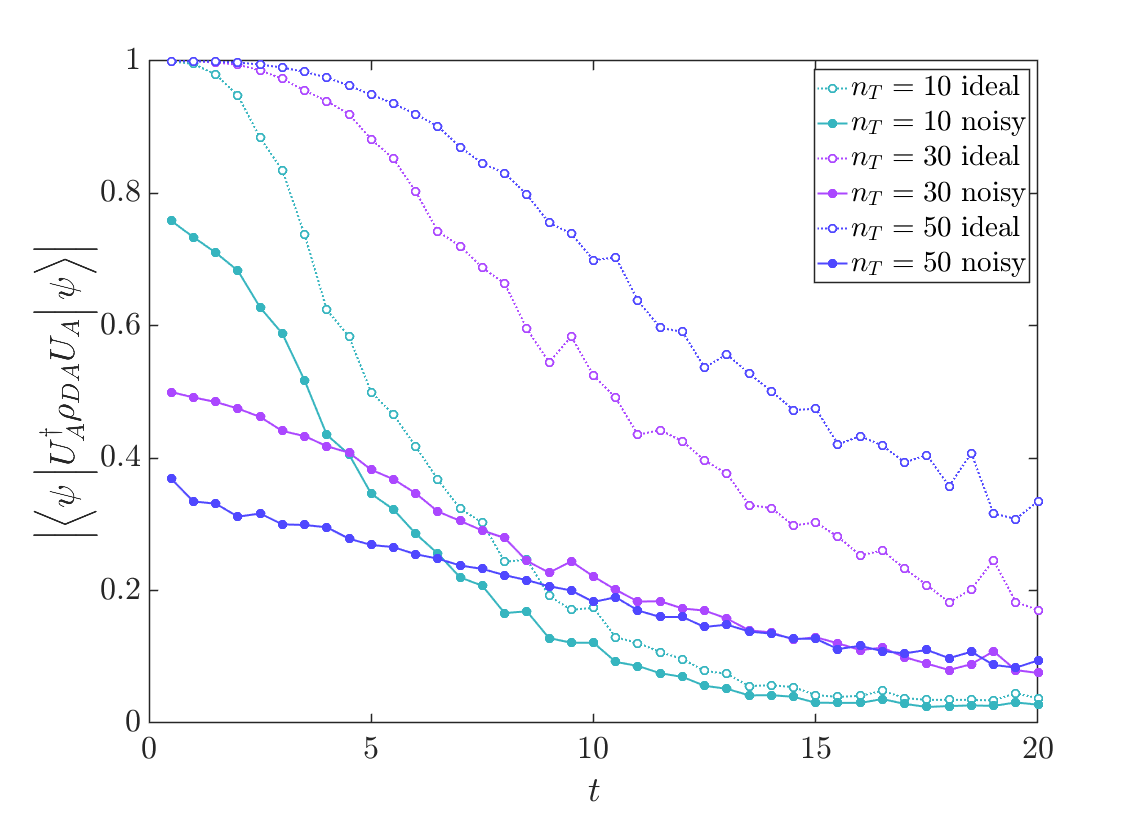}
    b)
    \includegraphics[width=\linewidth,clip,trim={1.1cm 0.4cm 3.3cm 1.4cm}]{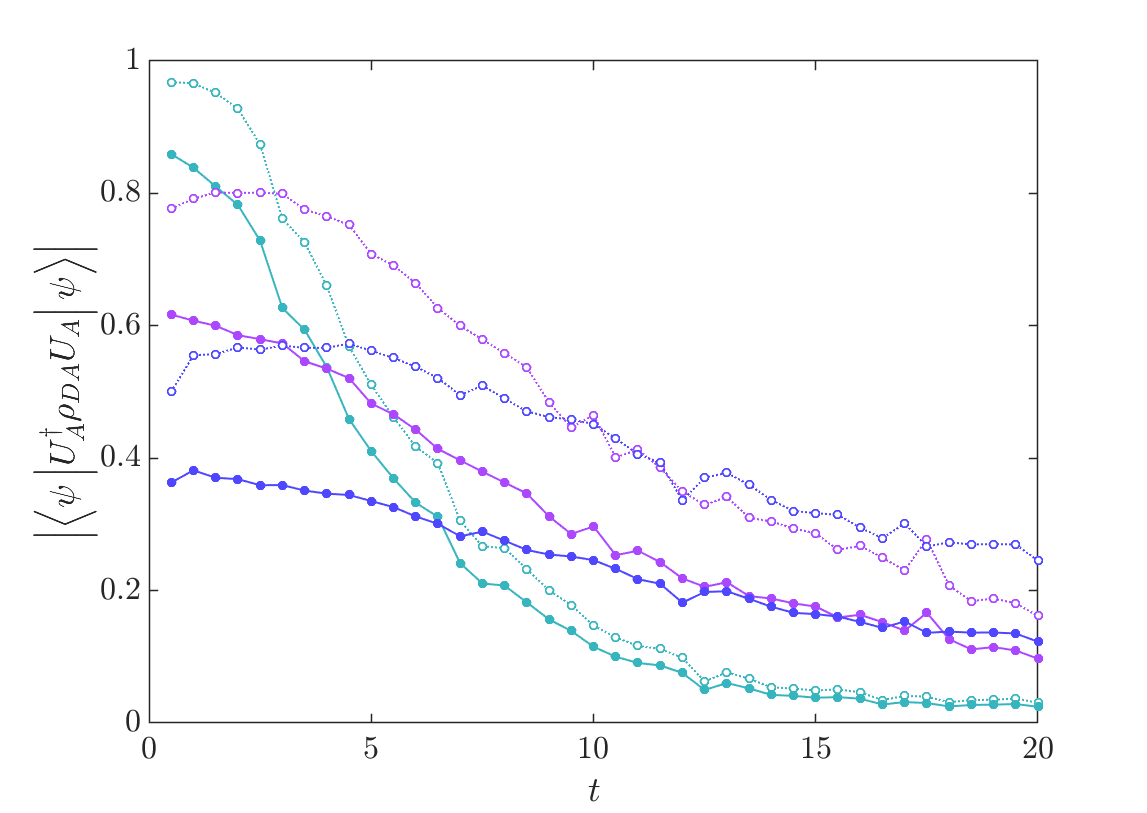}
    \caption{\textbf{Fidelity for the ladder architecture}. Fidelities for different number of Trotter steps and times for the (a) stepwise and (b) banged DA circuits. Here we employ $\alpha=\gamma=1$. The rest of the parameters are the same as in Fig.\ref{fig:snakeFidelity}. In this architecture, the digital analog can be implemented faster than in the NN architecture. As a consequence, if we maintain the time for applying a SQG, the effect of the decoherence and the dephasing affect less the outcome of the experiments. This results in an improvement on the fidelity, specially for the bDAQC circuits.}
    \label{fig:ladderFidelity}
\end{figure}

\begin{figure}
    \raggedright
    a)
    \includegraphics[width=\linewidth,clip,trim={1.1cm 0.4cm 3.3cm 1.4cm}]{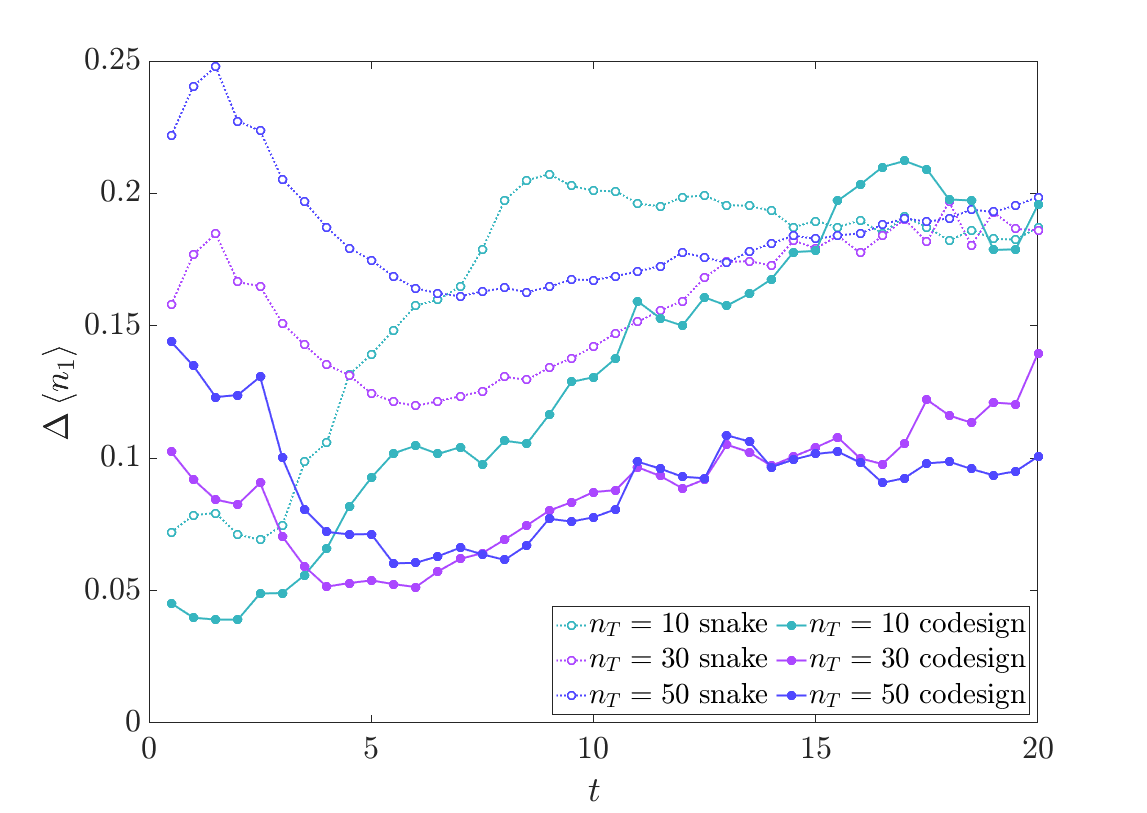}
    b)
    \includegraphics[width=\linewidth,clip,trim={1.1cm 0.4cm 3.3cm 1.4cm}]{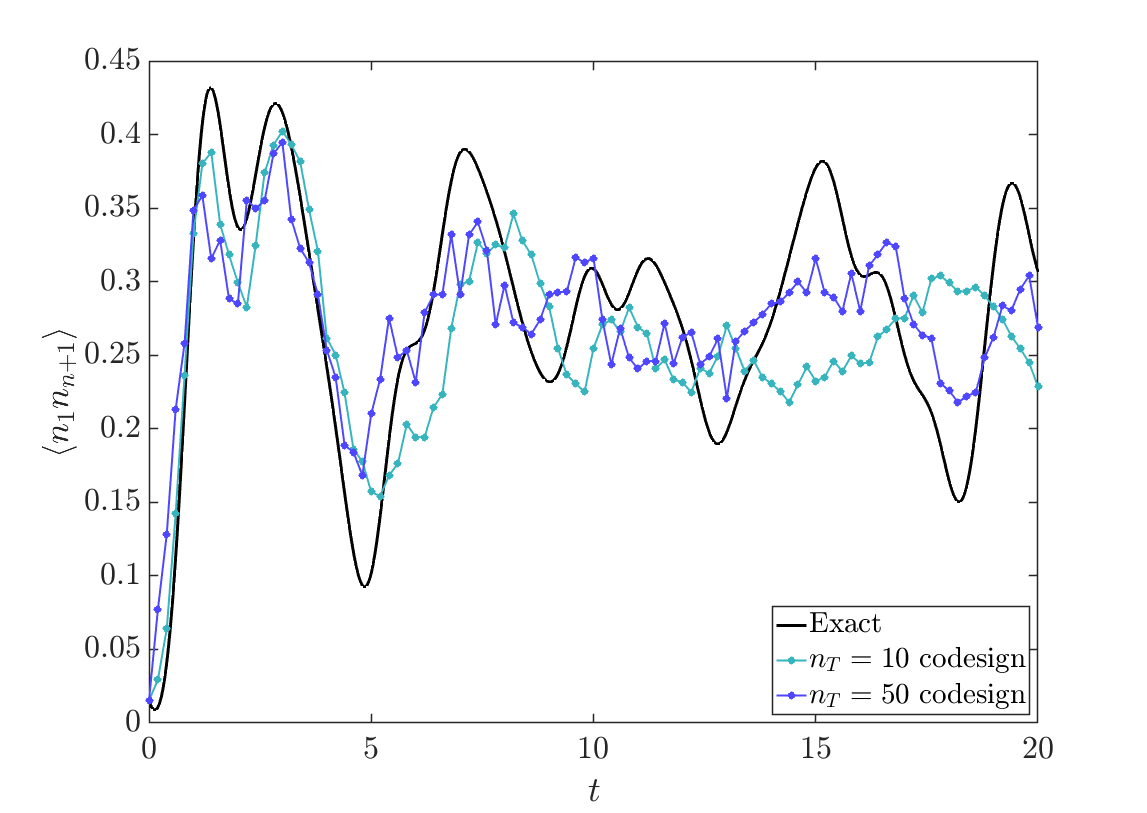}
    \caption{\textbf{Errors in measuring observables}. (a) Difference between the expected value and the simulated measurements of the total density for the first site. In this plot, we show the simulations of the noisy bDAQC circuits for the two topologies, with the same setup as in Fig.\ref{fig:snakeFidelity}. As expected from the results of the fidelity, we obtain a more accurate measurement of the expected value of the observable for the case in which we employ the codesigned hardware. We notice that there is an asymptotic behaviour on the error of measuring the observables, close to the 0.2 mark, as seen for the ``snake" topology and the codesigned experiment with $n_\text{T}=10$. Although not shown, the asymptotic behaviour is also present for the on-site double-occupancy observable. (b) Measurements for the on-site double-occupancy of the first site for different times and for the same initial state for the noisy bDAQC circuit in the codesigned hardware. In this case, we see that the simulation with $n_\text{T}=50$ yields a closer expected value to the exact one for times closer to $t=20$ than the simulation with $n_\text{T}=50$.}
    \label{fig:observableError}
\end{figure}

In Fig.~\ref{fig:snakeFidelity} and Fig.~\ref{fig:ladderFidelity} we show the average fidelity over a number of initial states for the sDAQC and bDAQC with and without noise. For simplicity, we call ``ideal" to the simulations in which we omit the noise. The number of runs is set so that the error in the fidelity is less than 0.001, with a maximum of 4000 runs per data point. Figure~\ref{fig:observableError} also shows the results of the circuits when measuring the observables from Eqs.~\eqref{eq:density} and~\ref{eq:correlations}. The main conclusion we extract from these studies is the confirmation that the fidelity for simulating the evolution over longer times drops with the number of Trotter steps. If we compare the stepwise and the banged implementations we see the expected results shown in Ref.~\cite{Parra-Rodriguez2020}, where we see a drop in the fidelity due to the unwanted interactions during the application of the SQGs in bDAQC. Furthermore, we see that for lower times, the fidelity in bDAQC decreases as we increase the number of Trotter steps. This effect comes from the reduction of the length of the analog blocks, which weakens the assumption that the time to apply a SQG is negligible compared to the analog block times. Therefore, there is an optimum number of Trotter steps which maximizes the fidelity for each simulation time $t$ in bDAQC.

When considering the errors, there is a considerable drop in fidelity, specially for longer simulation times. The simulations in the linear device shows that for the circuits with a greater number of Trotter steps, the fidelity is almost comparable to the one of a thermal state. The main contribution to this increase in noise is the overhead in the analog blocks required to implement the SWAP gates. On the contrary, for the ladder architecture there are no SWAP gates, and thus, the total simulation time stays close to constant. This allows us to increase the number of Trotter steps while only increasing the banging error.

When we employ a codesigned circuit to simulate the fermionic dynamics, the total time for the analog blocks depends solely on the time for the simulation and the relation between the coupling strength of the hardware and the corresponding couplings we want to simulate. Unlike the simulation with a NN hardware where we need to apply a number of SWAP gates that scales linearly with the number of Trotter steps, the simulation on the codesigned hardware is SWAP-free. An unwanted consequence of this is that for shorter simulation times, the bDAQC implementation on the codesigned hardware yields lower fidelities compared to the bDAQC implementation on the NN hardware. Assuming that we can tune the strength of the couplings before the experiments at will, we can avoid this problem by setting their strength such that the time of the shorter analog block is much grater than the time to apply a SQG, $\text{min} T_\text{analog}\gg T_\text{SQG}$.

\subsection{Simulation of noisy circuits}
\label{sec:error}

For the numerical experiments, we have simulated both the ideal (noiseless) and the noisy circuits. For modeling the errors in the DAQC circuits, we have followed the procedure proposed in Ref.~\cite{Paula2021}. There, they introduce 4 main sources of error.

The bit-flip error models the random switch of the state of a qubit. This error can be triggered by the control signals or by random thermal fluctuations. The simulation of this error is made by the quantum channel formalism, with a probability of a bit-flip to happen within a digital or analog block of $p_\text{bf}$. 

Both the decoherence and the dephasing of the qubits are modeled by the generalized damping channel and the dephasing channel respectively. Here, we assume that the state of the qubit can be de-exited due to the contact with an environment, with a thermal population of the ground state $p_\text{th}$. The rate at which these error occurs increases with the duration $t$ of the process, $\exp(-t/T)$ with $T_1$ the thermal relaxation time for the longitudinal relaxation and $T_2$ the phase coherence time.   

Additionally, they considered the error of applying a SQG. This error can account for the imperfect control pulses to perform the SQGs or from a magnetic field affecting the qubits. For this, they introduced a random deviation from the ideal unitary. In this paper, we simulated this error by introducing a error $\varepsilon$ with a normal distribution centered at 0 with standard deviation $r_\text{D}$, such that $\exp(-i\theta\mu)\rightarrow\exp(-i(\theta+\varepsilon)\mu)$, with $\mu$ the corresponding Pauli matrix. 

The last source of error comes from the imperfect control over the time at which the gates are applied. This instrumental error is modeled as a normal distribution of standard deviation $r_\text{B}t_\text{SQG}$ around the the ideal time at which each analog or digital block should start. As an example, an analog block applied between times $t_i$ and $t_{i+1}$ has is simulated as $\exp(-i (t_{i+1}-t_i)H_\text{I})\rightarrow\exp(-i (t_{i+1}-t_i+\delta)H_\text{I})$.

\section{Conclusions}

In summary, we have presented a digital-analog quantum algorithm for the quantum computation of the paradigmatic 1D Hubbard model. For the implementation, we have proposed two realistic circuit topologies, a 1D nearest-neighbour system and a codesigned system with a ladder architecture, in both cases with an Ising Hamiltonian as a resource. For the case of the 1D system, we have found a mapping which reduces drastically the number of SWAP gates, which grows linearly with the number of Trotter steps and it is independent of the number of fermions in the system. By inspecting the fidelity, and the evolution of the density and double-occupancy operators, we have shown that the algorithm has a very good performance, specially in the codesigned case. The performance of this algorithm is bounded by the errors coming from the Trotterization, which we have reduced by employing a symmetric expansion with negligible extra cost. We have run the experiments in both an ideal scenario for bounding the fidelity of the algorithm, and a realistic noisy scenario. The results obtained shows the validity of our algorithm for simulating the system using the DAQC protocol, specially when employing the bDAQC protocol in the codesigned architecture. When applied to a specific quantum platform, more detailed considerations on error sources and decoherence issues will be required, but evidences based on previous works show the better scalability of digital-analog than of purely digital approaches~\cite{Parra-Rodriguez2020,Martin2020}. This work is an important step in the development of quantum algorithms using DAQC techniques.\\

\noindent
\textbf{Acknowledgments.--} The authors acknowledge financial support from Basque Government QUANTEK project from ELKARTEK program (KK-2021/00070), Spanish Ramón y Cajal Grant RYC-2020-030503-I and the project grant PID2021-125823NA-I00 funded by MCIN/AEI/10.13039/501100011033 and by ``ERDF A way of making Europe” and ``ERDF Invest in your Future”, as well as from QMiCS (820505) and OpenSuperQ (820363) of the EU Flagship on Quantum Technologies, and the EU FET-Open projects Quromorphic (828826) and EPIQUS (899368). LCC also thanks the Brazilian Agencies CNPq, FAPEG and the Brazilian National Institute of Science and Technology of Quantum Information (INCT/IQ). This study was financed in part by the Coordena\c{c}\~ao de Aperfei\c{c}oamento de Pessoal de N\'ivel Superior - Brasil (CAPES) - Finance Code 001. MGdA acknowledges support from the UPV/EHU and TECNALIA 2021 PIF contract call.


\end{document}